\documentclass[10pt,letterpaper,conference]{IEEEtran}
\IEEEoverridecommandlockouts
\usepackage{cite}
\usepackage{amsmath,amssymb,amsfonts}
\usepackage{algorithmic}
\usepackage{graphicx}
\usepackage{textcomp}
\usepackage{xcolor}
\usepackage{subcaption}
\usepackage{tcolorbox}
\usepackage{etoolbox}
\def\BibTeX{{\rm B\kern-.05em{\sc i\kern-.025em b}\kern-.08em
    T\kern-.1667em\lower.7ex\hbox{E}\kern-.125emX}}
    
\newtoggle{SRCSUBMISSION}
\toggletrue{SRCSUBMISSION}
\begin{document}

\title{Micro BTB: A High Performance and Lightweight Last-Level Branch Target Buffer for Servers
}

\iftoggle{SRCSUBMISSION}{
\author{\IEEEauthorblockN{Vishal Gupta}
\IEEEauthorblockA{\textit{Department of Computer Science and Engineering} \\
\textit{Indian Institute of Technology, Kanpur}\\
vishal@cse.iitk.ac.in}
\and
\IEEEauthorblockN{Biswabandan Panda}
\IEEEauthorblockA{\textit{Department of Computer Science and Engineering} \\
\textit{Indian Institute of Technology, Bombay}\\
biswa@cse.iitb.ac.in}
}}{}

\maketitle

\begin{abstract}
High-performance branch target buffers (BTBs) and the L1I cache are key to high-performance front-end. Modern branch predictors are highly accurate, but with an increase in code footprint in modern-day server workloads, BTB and L1I misses are still frequent. Recent industry trend shows usage of large BTBs (100s of KB per core) that provide performance closer to the ideal BTB along with a decoupled front-end that provides efficient fetch-directed L1I instruction prefetching. On the other hand, techniques proposed by academia, like BTB prefetching and using retire order stream for learning, fail to provide significant performance with modern-day processor cores that are deeper and wider. 
\\ \indent We solve the problem fundamentally by increasing the storage density of the last-level BTB. We observe that not all branch instructions require a full branch target address. Instead, we can store the branch target as a branch offset, relative to the branch instruction. Using branch offset enables the BTB to store multiple branches per entry. We reduce the BTB storage in half, but we observe that it increases skewness in the BTB. We propose a skewed indexed and compressed last-level BTB design called MicroBTB (MBTB) that stores multiple branches per BTB entry. We evaluate MBTB on 100 industry-provided server workloads. A 4K-entry MBTB provides 17.61\% performance improvement compared to an 8K-entry baseline BTB design with a storage savings of 47.5KB per core.   

\end{abstract}


\section{Introduction}

Branch predictor and branch target buffer (BTB) are the two key front-end structures that play a significant role in providing high performance. Branch predictors predicts the direction (taken or not-taken) of a conditional branch. BTB provides information about whether an instruction is a branch instruction or not, and the target of the branch instruction. To prevent the processor from going on the wrong execution path, both the structures need to be accurate. Modern branch predictors like hashed perceptron\cite{perceptron} and TAGE\cite{tage} are highly accurate. However, with the increase in code footprint in server workloads, BTB and L1I cache see frequent misses. The code footprint ranges in multi-megabytes because of the deepening software stack with frequent updates/patches. 

\begin{figure}[t]
\centerline{\includegraphics[width=\linewidth]{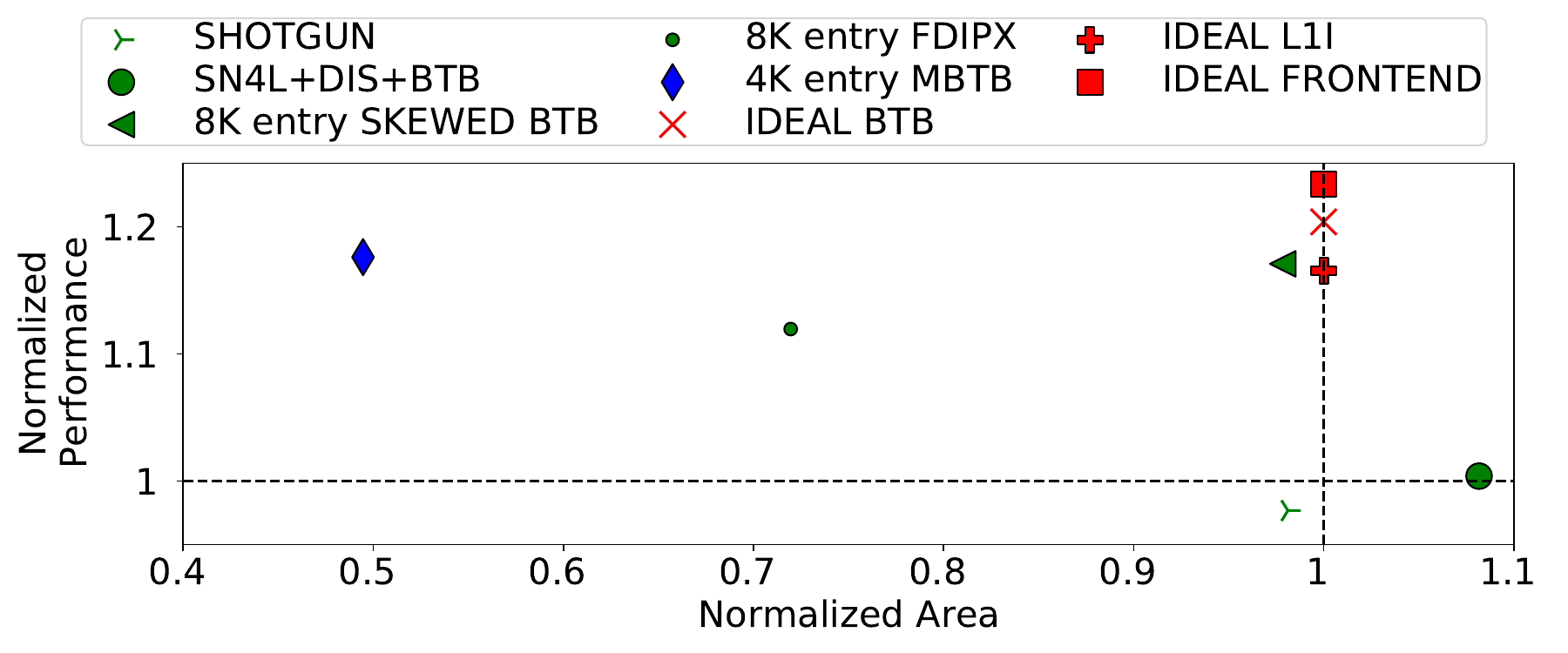}}
\vspace{-0.1in}
\caption{Performance improvements of state-of-the-art BTB designs scaled to 8K entry, normalized to the baseline system with 8K entry BTB (Table~\ref{table:parameters}) for 100 industry provided server workloads. Area overheads (per core) are normalized to 8K entry baseline BTB.} 
\label{fig:motivation}
\iftoggle{SRCSUBMISSION}{\vspace{-0.2in}}{
\vspace{-0.3in}}
\end{figure}

State-of-the-art techniques for reducing front-end bottlenecks like Shotgun\cite{shotgun} and SN4L+Dis+BTB\cite{divide} employ L1I prefetcher to prefetch cache blocks in the L1I. These techniques decode the prefetched cache blocks and pre-fill the branches into the BTB. With an increase in code footprint, L1I misses are high, and the L1I prefetchers in these techniques are not able to prefetch the instruction blocks in a timely manner. L1I prefetcher in Shotgun learns the control flow of a program using the retire order instruction stream. With modern-day high-performing processors becoming deeper and wider with each generation, the fetch stage of the pipeline is now much further ahead in the instruction stream compared to the retire stage of the pipeline. We show in Section \ref{section:evaluation} that the increase in the number of in-flight instructions cause lateness in Shotgun, resulting in lower performance. 

Recent industry trend shows that modern processors employ a decoupled front-end \cite{arm,amd,samsung,ibmz15} with a multi-level BTB design. A decoupled front-end decouples the branch prediction unit from the instruction fetch and allows the branch predictor to generate the address of future instructions. Modern processors like Arm Neoverse N1\cite{arm}, AMD Zen2 \cite{amd}, Samsung Exynos M3 \cite{samsung} and IBM z15 \cite{ibmz15} have high BTB capacities of 6K, 7K, 16K and 144K entries, respectively.  With technology scaling slowing down \cite{tech_scaling}, the number of transistors available because of reduction in transistor size is slowing down. BTB size cannot increase with an increase in code footprint without reducing the size of other structures, which creates a need to reduce the size of the BTB while providing similar or better performance. 

Previous work\cite{indirection} for reducing the storage overhead of BTB uses a Page Number (PN) cache to store the page numbers and use a smaller index to the PN-cache instead of a page number, for storing branch tag and target. The PN-cache needs to be accessed when accessing the BTB and also to get the branch target if it is a BTB hit. The PN-cache adds extra latency to the critical path of accessing the BTB.

\noindent \textbf{Opportunity:} Figure \ref{fig:motivation} shows average performance improvement with various state-of-the-art BTB designs \cite{fdipx,shotgun,divide,skewseznec} normalized to a baseline system\cite{sunny} with two-level BTB, a 128-entry L1 BTB and an 8192-entry L2 BTB. We do not show the performance and area overheads for Phantom BTB\cite{shift, phantom} and Confluence \cite{confluence} for 8K-entry BTB as it demands significant storage. Our baseline front-end is a decoupled one with a 24-entry (192 instructions) fetch target queue (FTQ) and a fetch directed instruction prefetcher (FDIP)\cite{fdip}. We use a 57-bit virtual address (keeping in mind the recent trend of a five-level page table\cite{fiveptw}) with a 57-bit instruction pointer (IP). We use 100 industry provided server workloads, which are available through the 1st championship value prediction (CVP-1) \cite{cvp}. The 1st instruction prefetching championship (IPC-1) \cite{ipc1} uses a subset of these traces for evaluation. We extend ChampSim \cite{champsim} with detailed extensions to the front-end and the memory hierarchy. In the baseline system (Table \ref{table:parameters}) for 100 server workloads, the average BTB and L1I MPKI are 8.6 and 54.94, respectively.

Figure \ref{fig:motivation} shows the performance improvement with the ideal BTB, ideal L1I, and the ideal front-end. For the ideal BTB, the first instance of a branch instruction is a miss, and the rest of the instances are hits. For the ideal L1I, L1I misses are converted to hits only when there is some space left in the miss status handling registers (MSHRs). This method of calculating ideal numbers for the L1I takes into the account the bandwidth constraint between L1I and L2. The ideal front-end has both ideal BTB and ideal L1I. 

A recent work \cite{fdipispass} has shown that with a decoupled front-end, an FDIP prefetcher, and the ideal BTB, the ideal L1I does not provide a significant performance improvement. We also observe the same in Figure \ref{fig:motivation} where an ideal front-end provides 3\% more improvement compared to the ideal BTB. Figure \ref{fig:motivation} also shows that state-of-the-art BTB designs like Shotgun and SN4L+Dis+BTB provide marginal performance improvement or require high storage like Skewed BTB\cite{skewseznec}. Please note that Figure \ref{fig:motivation} shows the performance of state-of-the-art BTB designs with a decoupled front-end. To the best of our knowledge, this is the first work that evaluates the impact of a decoupled front-end on state-of-the-art BTB designs. 

\noindent \textbf{Our Contributions:} We observe that not all branches require a full target, and we can encode the branch target in fewer bits. FDIP-X\cite{fdipx} uses four different BTBs to store branches with different encoding types. We propose a single compressed L2BTB design called MicroBTB (MBTB), storing one or multiple branches per BTB entry. MBTB uses skewed indexing to reduce the conflict misses arising due to halving the storage. 

\noindent In summary, we make the following key contributions:-
\begin{itemize}
    \item We show that the BTB misses are costly compared to the L1I misses with a decoupled front-end (Section \ref{section:motivation}).
    \item We propose a skewed and compressed L2BTB design called MicroBTB (MBTB) to mitigate these costly BTB misses (Section \ref{section:gangbtb}).
    \item We show that a 4K-entry MBTB provides an average performance improvement of 17.61\% compared to an 8K-entry baseline BTB, resulting in 51\% storage savings (Section \ref{section:evaluation}).
\end{itemize}

\begin{figure*}[t]
        \begin{subfigure}[b]{0.3\textwidth}
                \includegraphics[width=\linewidth]{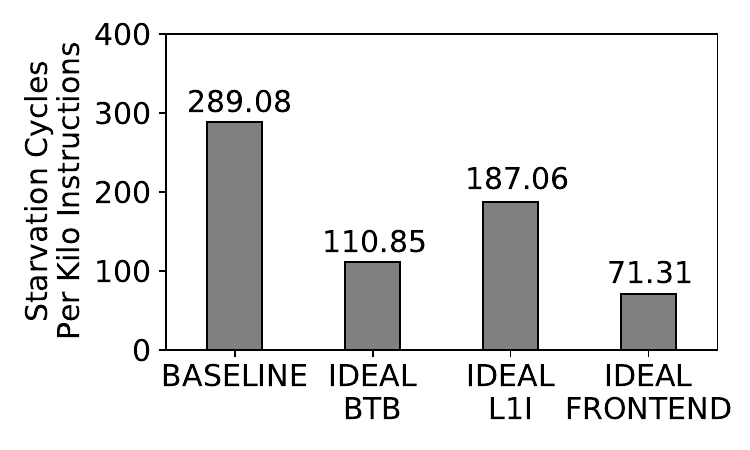}
                \caption{}
        \end{subfigure}%
        \hfill
        \begin{subfigure}[b]{0.7\textwidth}
                 \includegraphics[width=\linewidth]{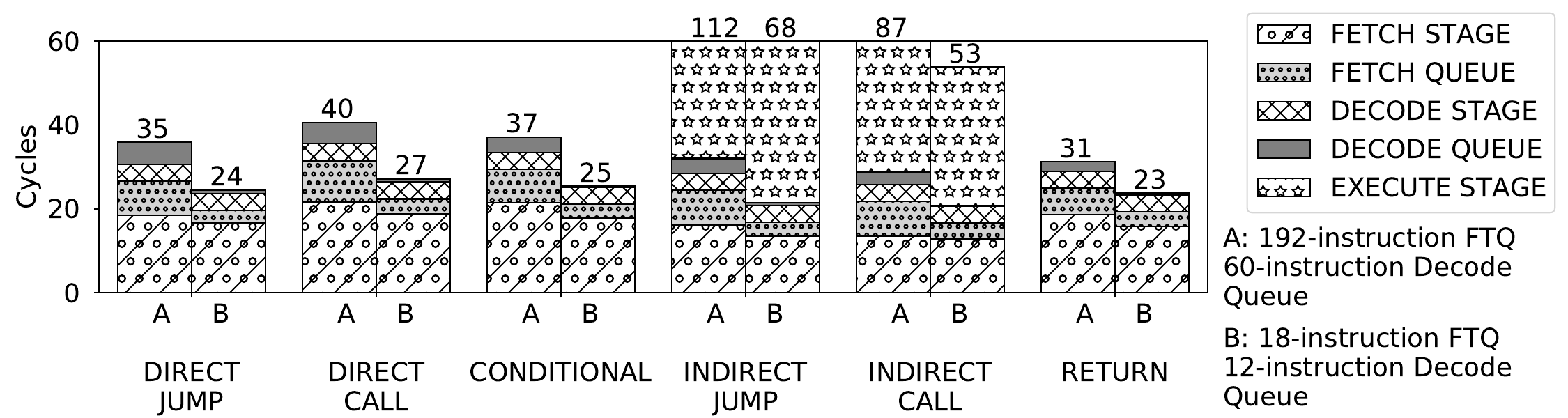}
                 \caption{}
        \end{subfigure}%
        \vspace{-0.1in}
\caption{(a) Starvation cycles per kilo instructions for baseline, ideal BTB, ideal L1I, and the ideal front-end (b) Breakdown of branch resolution latency with front-ends having two different queue sizes for the frontend structures: FTQ and Decode queue.}\label{fig:ideal_stall}
\iftoggle{SRCSUBMISSION}{\vspace{-0.2in}}{
\vspace{-0.2in}}
\end{figure*}
\section{Related Work}
\label{section:background}

In this section, we provide an overview of some of the state-of-the-art BTB designs. \\
\noindent \textbf{Phantom BTB\cite{phantom}:}  Phantom BTB proposes a hierarchical BTB design where at the first level, it uses a conventional BTB. However, it exploits temporal correlation among control flow jumps at the second level, packs multiple first-level BTB misses as temporal groups, and stores them at the last-level cache (LLC) using virtualization. A second-level BTB with 4K temporal groups can demand up to 256KB of LLC space shared among all the cores. Unfortunately, with this design, the performance does not improve significantly as the level-1 BTB incurs frequent misses, and getting a response from the level-2 BTB is usually delayed because it is limited by the LLC access latency. \\
\noindent \textbf{Air BTB\cite{confluence}:} Air BTB is a block-based BTB design in which the BTB is in sync with the L1I cache. It stores a branch bitmap to identify branch instructions inside the cache block. Prefetch or demand blocks which are filled into the L1I cache are predecoded, and the branches are stored in the Air BTB. \\
\noindent \textbf{Shotgun BTB\cite{shotgun}:} Shotgun BTB proposes a new BTB design segregated by the type of control flow jumps. It uses three kinds of BTB: (i) C-BTB for conditional branches, (ii) U-BTB for unconditional branches, and (iii) RIB for return. Shotgun insight is that the global control flow of a program is determined by unconditional branches, whereas conditional branches determine the local control flow. Most of the BTB is dedicated to unconditional branches where it stores the spatial footprint and uses that for prefetching the next instruction blocks. Shotgun pre-decodes conditional branches from these prefetched blocks, which helps achieve a high hit rate for conditional branches despite its small size. \\
\noindent \textbf{SN4L+Dis+BTB \cite{divide}:} SN4L+Dis+BTB uses baseline BTB design but uses an additional BTB prefetch buffer to hold predecoded branches. It proposes a next-line and discontinuity-based prefetcher for L1I prefetching and performs a Shotgun style BTB prefetching by pre-decoding the prefetched blocks. \\
\noindent \textbf{FDIP-X \cite{fdipx}:} FDIP-X uses four BTBs with different branch target offsets, which is the distance between the branch instruction and its target. The insight that drives FDIP-X is that branch offset lengths are not distributed equally: conditional branches have shorter offsets than unconditional branches, enabling FDIP-X to store a higher number of branches in the same storage budget. \\
\noindent \textbf{Skewed BTB\cite{skewseznec}:} Skewed BTB design uses different set indexing function for each way to increase the utilization of BTB entries. Utilization increases because a branch instruction that can go to a single set in the baseline design can go to multiple sets with the skewed BTB design.  \\
\begin{figure}[t]
\centerline{\includegraphics[width=\linewidth]{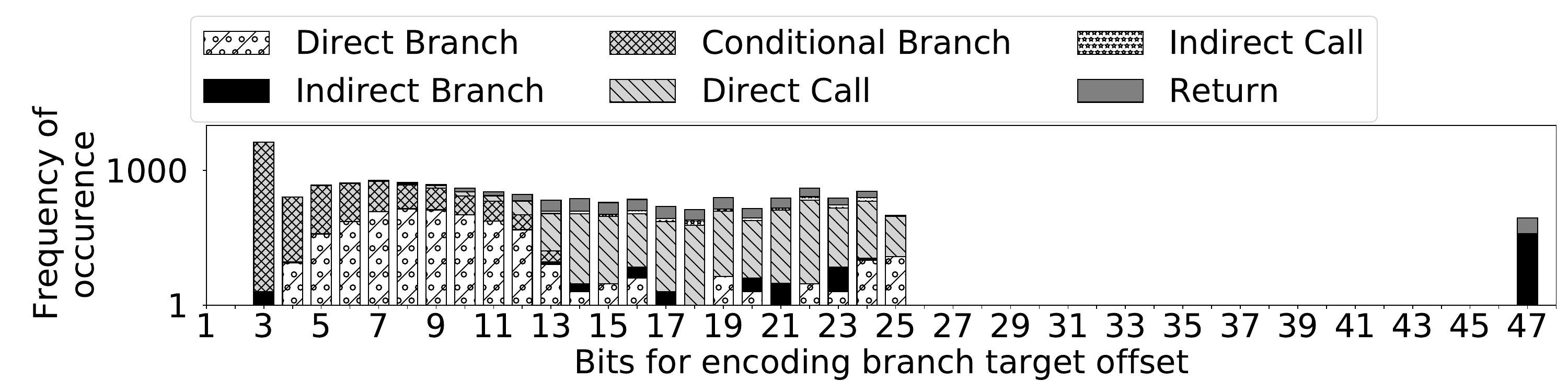}}
\vspace{-0.1in}
\caption{Frequency of occurrence of branch instructions whose target is within a specified offset. Offset refers to the distance between branch IP and its target.}
\label{fig:offset_freq}
\iftoggle{SRCSUBMISSION}{\vspace{-0.2in}}{
\vspace{-0.22in}}
\end{figure}

\section{Motivation}
\label{section:motivation}

In this section, we first elaborate on why BTB misses are costly compared to the L1I misses. We then show that the BTB miss cost increases with an increase in queue size of the front-end structures like FTQ and decode queue. Finally, we discuss about how to improve the storage density of the BTB to reduce these costly BTB misses.

\begin{figure*}[t]
\centerline{\includegraphics[width=\linewidth]{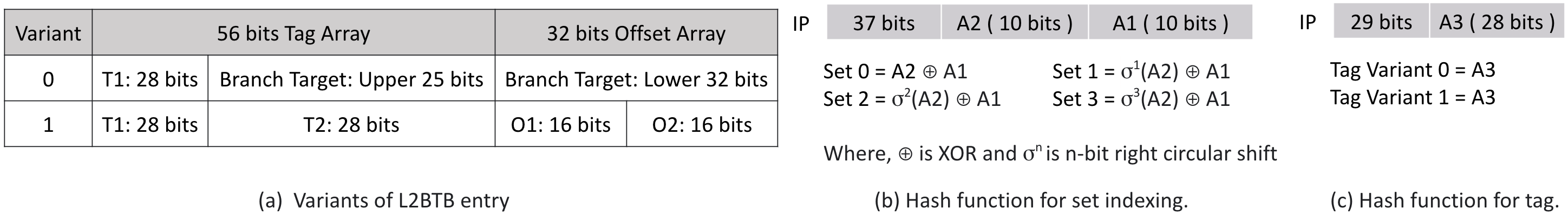}}
\caption{(a) Variants of MBTB entry. T1 and T2 store the hash of the branch instruction and O1 and O2 store the branch offset (b) Skewed function that generates four indices for four banks (c) Tag calculation for different MBTB entry variants.}
\label{fig:l2btb_entry}
\iftoggle{SRCSUBMISSION}{\vspace{-0.0in}}{
\vspace{-0.1in}}
\end{figure*}

\begin{figure*}[t]
        \begin{subfigure}[b]{0.65\textwidth}
                \includegraphics[width=\linewidth]{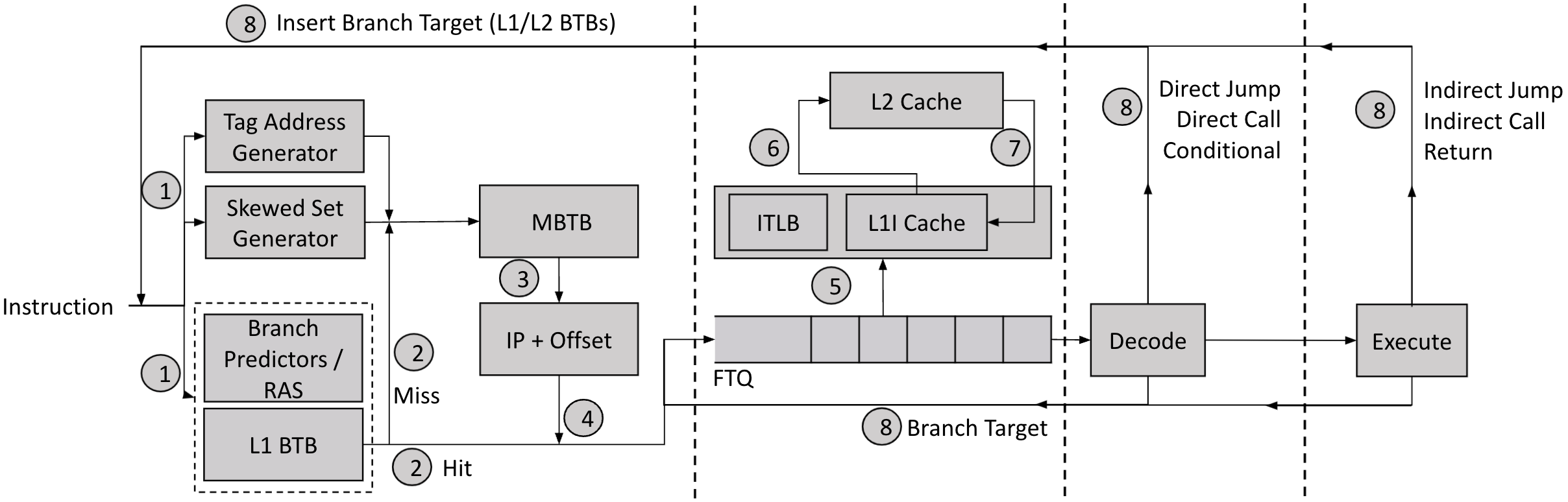}
                \caption{}
        \end{subfigure}%
        \hfill
        \begin{subfigure}[b]{0.33\textwidth}
                 \includegraphics[width=\linewidth]{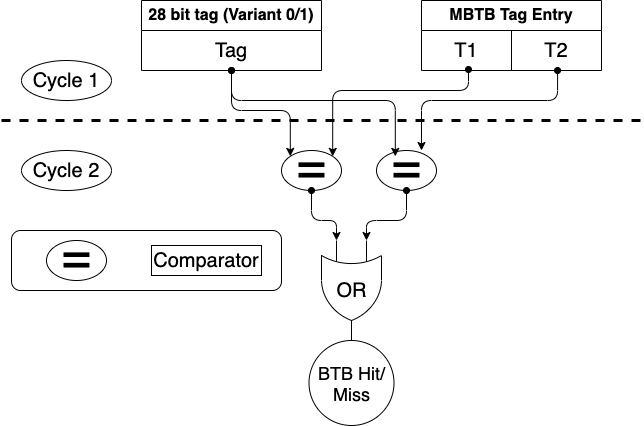}
                \caption{}
        \end{subfigure}%
\vspace{-0.1in}
\caption{(a) Flow of a branch instruction with a decoupled front-end and the MBTB (b) Implementation of the MBTB tag comparison.}
\label{fig:idea}
\iftoggle{SRCSUBMISSION}{\vspace{-0.1in}}{
\vspace{-0.22in}}
\end{figure*}

\noindent \textbf{BTB misses are costly compared to L1I misses:} Figure \ref{fig:motivation} shows that the ideal front-end provides only 3\% performance improvement on top of the ideal BTB. To understand this performance trend, we use the metric \emph{starvation cycles per kilo instructions (SCKI)}. This metric provides the number of cycles for which the decode stage is empty, and no instruction is sent to the ROB. Higher SCKI indicates that the front-end is a bottleneck as it under-utilizes the bandwidth between decode stage and ROB. Figure~\ref{fig:ideal_stall}(a) shows that with an ideal BTB, the SCKI reduces to 110 from 289, but with an ideal front-end, the SCKI reduces marginally to 71. 

On top of ideal BTB, ideal L1I does not decrease the SCKI, significantly, because with a decoupled front-end, on an L1I miss, the branch predictor continues to predict future instructions. The request for cache block containing these instructions is sent to the L1I, effectively hiding the latency for future instructions as multiple L1I requests are going in parallel. Recent work\cite{fdipispass} also shows that L1I prefetchers like EIP\cite{eip} and FNL+MMA\cite{fnlmma} do not provide significant performance improvement with a decoupled front-end and FDIP prefetcher.

On a BTB miss, the fetch pipeline is either stalled if the branch predictor predicts the direction of the branch is taken with high confidence or the pipeline goes on the wrong path until the branch instruction executes. For a direct branch/call instruction, the branch gets resolved in the decode stage, whereas for indirect branch/call instruction, the branch gets resolved in the execute stage. Once the branch instruction is resolved, the front-end is flushed if it was going on the wrong path, and the fetch stage continues with the correct branch target.

After resolving the branch, the front-end is re-steered to the correct path. There are multiple cycles for which the decode stage is empty, and nothing is added to the ROB, while the instructions gets fetched from the L1I cache. The back-end of the processor does not have any instructions to execute, making the BTB misses costly in terms of performance. This motivates us to design a BTB that can reduce the BTB misses and help in increasing the overall performance. 

Modern processors use a decoupled front-end with an increase in queue size for the front-end structures like FTQ to support high-performing FDIP prefetcher. We next discuss the impact of this increase in queue size on branch resolution latency.

\noindent \textbf{Impact of increasing queue size:} Figure~\ref{fig:ideal_stall}(b) shows the breakdown of branch resolution latency per branch type for a front-end with a large FTQ and a small FTQ. Large FTQ has 24 entries, where each entry can hold up to eight instructions from the same cache line, storing 192 instructions \cite{sunny}. Small FTQ has 18 entries, where each entry can store one instruction. This showcases the impact of different queue sizes on the branch resolution latency, which is the time it takes for a branch instruction to resolve a BTB miss. A branch instruction is first fetched from the L1I cache and then sent to the decode queue for decoding. 


On a BTB miss, for a return instruction, the front-end gets to know the type of instruction in the decode stage, and the fetch stage is re-steered to access the return address stack (RAS). Since the RAS accurately captures the return instruction target, the branch gets resolved after the decode stage. If the target stored in RAS is wrong, then the branch gets resolved in the execute stage. 

Figure~\ref{fig:ideal_stall}(b) shows that increasing the FTQ and decode queue size increases the branch resolution latency. The breakdown shows that the time spent in the fetch and decode queue increases as the size of the queue increases. The decode stage decodes a fixed number of instructions per cycle, but if the queue size increases, then the time spent by an instruction in the queue increases. This adds extra latency in resolving the branch, which forces the front-end to be on the wrong path for higher number of cycles. With a processor similar to our baseline system having a 24-entry FTQ (192 instructions) and a 60-instruction decode queue, there are many instructions in the front-end that need to be flushed once the mis-predicted branch is resolved, and the fetch stage moves to the correct target. For indirect branches, the penalty is more because the increase in queue size increases the number of instructions in the ROB. There is a delay in the execution of branch instruction because multiple instructions competes for the shared resource in the execute stage. 


\noindent \textbf{Scope for BTB Compression:} We alleviate this increase in BTB cost by increasing the storage density of the last-level BTB that requires storing multiple branches per BTB entry. Figure~\ref{fig:offset_freq} shows the branch offset bits required to encode the branch target of a branch instruction relative to the IP of the branch instruction. We find that storing a 57-bit branch target is not required for many branch instructions as observed in the previous work \cite{fdipx}. We store a branch IP and its offset in a BTB entry instead of the branch target. On a BTB hit, the offset is added to the branch IP to get the branch target. Since the bits required to store the offset are less than the branch target, a single BTB entry can store multiple branches. 
\\ \indent Figure~\ref{fig:offset_freq} also shows the branch target offset required per branch type. Conditional and direct jump instructions have shorter offsets as these branches define the local control flow of the program, whereas call/return instructions have longer offsets as these branches define the global control flow of the program \cite{shotgun}. 
\section{Micro BTB: Design and implementation}
\label{section:gangbtb}

Using our observation that multiple branches can be stored using offsets, we propose a skewed and compressed L2 BTB called MicroBTB (MBTB). MBTB stores a hash of the branch instruction as tag and branch target as offset. Our baseline front-end is a decoupled one with FTQ that stores the predictions from the branch prediction unit until these instructions get consumed by the L1I. L1BTB design is the same as in the baseline system. Halving the storage increase conflict misses, and we use skewed indexing to reduce these conflict misses with the MBTB. We use a return address stack (RAS) that is a part of the baseline front-end to get the branch target of a return branch instruction. We use ITTAGE \cite{tage} for indirect branch prediction. \\
\noindent \textbf{Compressed MBTB entry:} Figure~\ref{fig:l2btb_entry}(a) shows two different types of variants which an MBTB entry can store. We find that increasing the number of variants increases the complexity and decreases performance gains. Section \ref{section:evaluation} shows the effect of more variants with MBTB. Variant 0 is uncompressed, and it stores a single branch instruction and its target. Variant 1 stores two branch instructions per entry depending on the size of offsets bit required to encode the distance between branch instruction and its target. Each offset field's most significant bit stores the direction of the branch \textit{i.e.} whether it is a forward or a backward branch. Branch instruction that requires more than 15 bits for encoding the offset uses the uncompressed variant 0. Branch instructions that require less than 15 bits for encoding the offset uses the compressed variant. \\
\noindent \textbf{Skewed set and Tag address generator:} To minimize BTB conflict misses by spreading out the BTB accesses across all the sets, we generate four different MBTB set numbers using a skewing function\cite{skewbodin,skewedaddress}. The skewing function generates different set numbers mapped to four different banks for addresses that would have otherwise conflicted using the baseline indexing function. Figure~\ref{fig:l2btb_entry}(b) describes the skewing function used. Figure~\ref{fig:l2btb_entry}(c) shows that MBTB uses lower 28 bits of the branch IP for both variant 0 and variant 1.\\
\subsection{Design} \label{section:design}
Figure ~\ref{fig:idea}(a) shows the flow of a branch instruction through a decoupled front-end with MBTB. 
\\\noindent \textbf{BTB access:} The address of the branch instruction goes through L1 BTB and branch predictors (\raisebox{.5pt}{\textcircled{\raisebox{-.9pt} {1}}}). Simultaneously, the address goes through the skewed set generator and the tag address generator (\raisebox{.5pt}{\textcircled{\raisebox{-.9pt} {1}}}). If there is a hit in L1 BTB (\raisebox{.5pt}{\textcircled{\raisebox{-.9pt} {2}}}), then the branch target from L1 BTB goes to the FTQ and we drop the current request to MBTB.\\
\noindent If we get a miss in L1 BTB (\raisebox{.5pt}{\textcircled{\raisebox{-.9pt} {2}}}), then we access the MBTB. The skewed set generator provides four different set numbers for each bank. We check the variant and compare the tag present in each entry based on the variant, to check if it is a hit or not. Simultaneously, we extract the offset from MBTB entry. If there is a MBTB hit (\raisebox{.5pt}{\textcircled{\raisebox{-.9pt} {3}}}), then we add/subtract the offset to the branch IP based on the most significant bit in the offset field and send it to the FTQ (\raisebox{.5pt}{\textcircled{\raisebox{-.9pt} {4}}}). If there is a MBTB miss, then the branch instruction goes to the subsequent stages of the processor pipeline to get the branch target. \\
\noindent \textbf{Instruction fetch:} Instructions from the FTQ send requests to L1I (\raisebox{.5pt}{\textcircled{\raisebox{-.9pt} {5}}}). Since, L1I is virtually indexed, physically tagged (VIPT), ITLB is accessed to get the physical address and then L1I is accessed. If the request hits in L1I, the instruction is sent back to FTQ, else if it misses, it goes to the lower level of the cache hierarchy (\raisebox{.5pt}{\textcircled{\raisebox{-.9pt} {6}}}). Instruction cache block is fetched from the lower levels of cache hierarchy and filled in L1I (\raisebox{.5pt}{\textcircled{\raisebox{-.9pt} {7}}}) and the instruction is sent to the FTQ. \\
\noindent \textbf{BTB insertions and updates:} For direct jump/call and conditional types of branches, we get the target in the decode stage, and we insert the target into the L1 and L2 BTBs
(\raisebox{.5pt}{\textcircled{\raisebox{-.9pt} {8}}}). For indirect jump/call and return types of branches, we get the target after the branch is executed in the execute stage and then we update the L1 and L2 BTBs (\raisebox{.5pt}{\textcircled{\raisebox{-.9pt} {8}}}). Note that the FTQ also gets updated with the target. BTB updates during the wrong-path execution affect performance. However, with the MBTB, BTB MPKI is extremely low (more details in Section \ref{section:evaluation}) and the wrong-path BTB updates do not affect the overall performance improvement significantly. 

\subsection{Implementation} \label{section:implementation}

\textbf{MBTB} is a 4096-entry skewed L2BTB with four banks, each of 1024 sets that are direct-mapped, implemented as four SRAM tables. To get the effect of 4-way skewed associative mapping, we generate four BTB set numbers scattered across four banks for a given branch IP. On an MBTB insertion, if we do not find a BTB entry to allocate, then we choose a victim out of those four BTB entries through a random replacement policy. We find marginal utility with more sophisticated replacement policies as the BTB MPKI drops significantly due to compression. Each entry in MBTB has a 56-bit tag field, a 32-bit offset field, two bits to determine the branch type, and one bit to determine the variant, totaling 91 bits. \\
\noindent \textbf{Storage savings:} MBTB has 4096 entries where each entry is of 91 bits. The storage overhead for MBTB is \textbf{45.5KB}. The baseline system uses an 8192-entry L2BTB where each entry has a 32-bit tag field, a 57-bit target field, 2-bit LRU, and two bits to determine the branch type totaling 93 bits. The storage overhead for baseline L2BTB is \textbf{93KB}. MBTB provides a storage savings of \textbf{47.5KB} and an average performance improvement of \textbf{17.61\%} compared to an 8K entry baseline BTB. Section \ref{section:evaluation} provides detailed performance improvement.
\noindent \textbf{MBTB Tag Comparison:} The tag address and the skewed set generator for the MBTB operate on the \emph{same clock cycle} when the L1 BTB is accessed. On the first cycle, the skewed set generator generates four different set numbers mapped to four different banks from a 57-bit IP, and concurrently the tag address generator generates a tag for different variants. If there is an L1 BTB miss, then we check the MBTB. \\
\indent Figure ~\ref{fig:idea}(b) shows the steps for MBTB tag comparison. The tag field of the MBTB is of 56 bits, consisting of two 28-bit chunks. On cycle 1, the tags T1 and T2 are read from the MBTB entry selected by the skewed set generator. On cycle 2, the 28-bit comparator compares the tag bits from the tag address generator to the bits stored in the MBTB entry. The OR gate combines the result, and if the output is one, then it is a BTB hit, or else it is a BTB miss. All the offsets are extracted concurrently with the tag comparison process. After the tag comparison, the offset is added to the branch IP to get the branch target. The branch target is added to the FTQ. \\
\noindent \textbf{Storing Indirect Branches in MBTB:} Direct and conditional types of branches have a single target, for which the variant is fixed, unlike indirect branches. Indirect branches can have multiple offsets depending on the distance between branch IP and the target IP. An indirect branch stored in MBTB in variant-1 format can require variant-0 if the offset between the current IP and the branch target is more than 15 bits. In this case, we invalidate the entry with variant-1, and we select a victim amongst the four entries provided by the skewed set generator using a random replacement policy. We evict the victim entry and then insert the indirect branch with variant-0.\\
\noindent \textbf{Storing Return Branches in MBTB:} Return type of branch instructions do not need to store their branch target in the BTB as RAS is used to get the target. We still need to track the return instruction in the BTB because the pipeline only knows about the type of instruction in the decode stage. At the fetch stage, BTB helps in identifying branch instruction and its type. We store the return instructions with variant-1 as it takes the least storage.

\begin{table}[t]
\caption{Parameters of the baseline system.}
\centering
\scriptsize{
\begin{tabular}{|p{0.8cm}|p{7.4cm}|}
\hline
Processor  & 4 GHz, out of order, hashed perceptron branch predictor for conditional branches \cite{perceptron} that merges \cite{fast}, \cite{merge}, and \cite{perceptron} with branch prediction accuracy of 99.6\%. ITTAGE \cite{tage} for indirect branches, 24-entry FTQ, 60-entry decode queue, OoO width: 6, ROB size: 352, RAS: 32 entries \\\hline
BTB & L1BTB: 128 entry (64 sets, 2 way, LRU), 1 cycle \\
& L2BTB: 8192 entry (2048 sets, 4 way, LRU), 2 cycle \\\hline
Caches & L1I: 32KB (8-way), LRU, 16 MSHRs, FDIP prefetcher\cite{fdip}, 4 cycles \\
& L1D: 48KB (12-way), LRU, 16 MSHRs, IPCP prefetcher\cite{ipcp}, 5 cycles   \\
& L2: 512KB (8-way), SRRIP\cite{rrip}, 32 MSHRs, IPCP prefetcher\cite{ipcp}, 10 cycles\\
& L3: 2MB, 16 way, DRRIP\cite{rrip}, 64 MSHRs, 20 cycles \\\hline
\iftoggle{SRCSUBMISSION}{
TLBs & ITLB: 64-entry (4-way), LRU, 8 MSHRs, 1 cycle \\
& DTLB: 64-entry (4-way), LRU, 8 MSHRs, 1 cycle  \\
& STLB: 2048-entry (16-way), LRU, 16 MSHRs, 8 cycles \\\hline}{}
DRAM & 4 GB, 1 channel, 6400 MT/sec \\\hline
\end{tabular}}
\label{table:parameters}
\iftoggle{SRCSUBMISSION}{\vspace{-0.2in}}{
\vspace{-0.2in}}
\end{table}

\iftoggle{SRCSUBMISSION}{
\begin{table*}[t]
\caption{BTB designs with their storage requirements used in evaluation.}
\centering
\scriptsize{
\begin{tabular}{|p{1.4cm}|p{11.2cm}|p{2cm}|p{2cm}|}
\hline
Front-ends & BTB & L1I prefetcher & Total storage \\\hline
Baseline  &  2048 sets and 4 ways, 32-bit tag, 57-bit target IP, two LRU bits, two branch type bits & FDIP prefetcher & 93 KB  (8K entry) \\\hline
Shotgun & Basic block based BTBs: 4096 entry UBTB (4-way, 114 bits per entry), 2048 entry CBTB (4-way, 97 bits per entry), and 2048 entry RIB (4-way, 40 bits per entry) & BTB directed prefetcher & 91.25 KB (8K entry) \\\hline
SN4L+DIS+ BTB & Baseline BTB (93KB for 8K entry) & As described in the paper \cite{divide}, 7.6KB & 100.6 KB  (8K entry) \\\hline
Skewed BTB & 2048 sets and 4 way, 32-bit tag, 57-bit target IP, two branch type bits & FDIP prefetcher & 91 KB (8K entry) \\\hline
FDIPX BTB & 6144 entry 8-bit offset BTB (4 ways, 29 bits per entry), 6144 entry 13-bit offset BTB (4 way, 34 bits per entry), 6144 entry 23-bit offset (4 way, 44 bits per entry) and 896 entry 57-bit offset BTB (4 way, 77 bits per entry) & FDIP prefetcher & 66.92 KB \\\hline
MBTB & 1024 sets and 4 way, 56-bit tag field, 32-bit offset field, two branch type bits and one variant bits & FDIP prefetcher & \textbf{45.5 KB (4K entry)   91 KB (8K entry)}\\\hline
\end{tabular}}
\label{table:designs}
\iftoggle{SRCSUBMISSION}{\vspace{-0.2in}}{
\vspace{-0.22in}}
\end{table*}}{}

\iftoggle{SRCSUBMISSION}{
\begin{figure*}[t]
\centerline{\includegraphics[width=\linewidth]{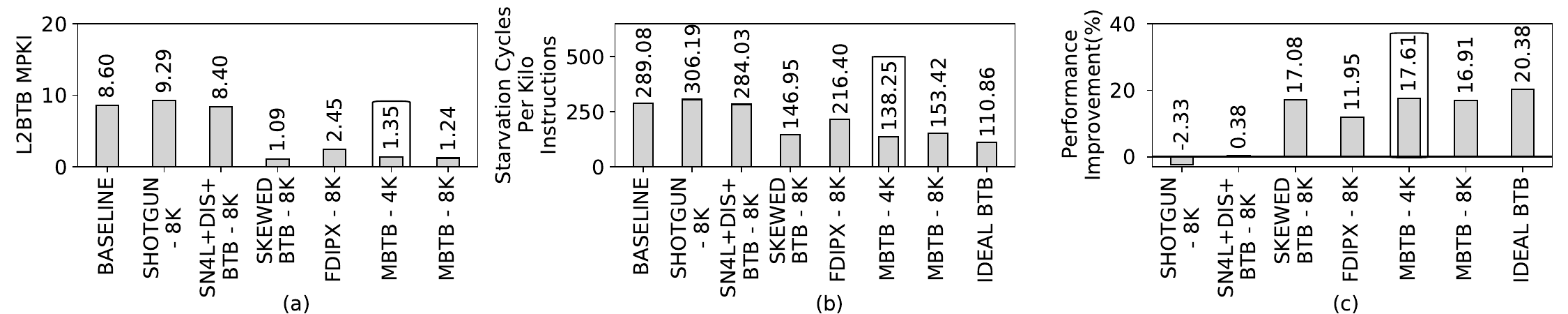}}
\caption{State-of-the-art BTB designs and MBTB evaluated on the following metrics :- (a) L2BTB Misses per Kilo Instructions (Lower the better) (b) Stall Cycles per Kilo Instructions (Lower the better) (c) Performance Improvement (Higher the better).} 
\label{fig:btb_mpki_sota}
\iftoggle{SRCSUBMISSION}{\vspace{-0.2in}}{
\vspace{-0.22in}}
\end{figure*}}{}
\section{Evaluation}\label{section:evaluation}

\subsection{Evaluation Methodology}

\iftoggle{SRCSUBMISSION}{}{
\begin{table*}[t]
\caption{BTB designs with their storage requirements.}
\centering
\scriptsize{
\begin{tabular}{|p{1.4cm}|p{11.2cm}|p{2cm}|p{2cm}|}
\hline
Front-ends & BTB & L1I prefetcher & Total storage \\\hline
Baseline  &  2048 sets and 4 ways, 32-bit tag, 57-bit target IP, two LRU bits, two branch type bits & FDIP prefetcher & 93 KB  (8K entry) \\\hline
Shotgun & Basic block based BTBs: 4096 entry UBTB (4-way, 114 bits per entry), 2048 entry CBTB (4-way, 97 bits per entry), and 2048 entry RIB (4-way, 40 bits per entry) & BTB directed prefetcher & 91.25 KB (8K entry) \\\hline
SN4L+DIS+ BTB & Baseline BTB (93KB for 8K entry) & As described in the paper \cite{divide}, 7.6KB & 100.6 KB  (8K entry) \\\hline
Skewed BTB & 2048 sets and 4 way, 32-bit tag, 57-bit target IP, two branch type bits & FDIP prefetcher & 91 KB (8K entry) \\\hline
FDIPX BTB & 6144 entry 8-bit offset BTB (4 ways, 29 bits per entry), 6144 entry 13-bit offset BTB (4 way, 34 bits per entry), 6144 entry 23-bit offset (4 way, 44 bits per entry) and 896 entry 57-bit offset BTB (4 way, 77 bits per entry) & FDIP prefetcher & 66.92 KB \\\hline
MBTB & 1024 sets and 4 way, 56-bit tag field, 32-bit offset field, two branch type bits and one variant bits & FDIP prefetcher & \textbf{45.5 KB (4K entry)   91 KB (8K entry)}\\\hline
\end{tabular}}
\label{table:designs}
\iftoggle{SRCSUBMISSION}{\vspace{-0.2in}}{
\vspace{-0.22in}}
\end{table*}}

\iftoggle{SRCSUBMISSION}{}{
\begin{figure*}[t]
\centerline{\includegraphics[width=\linewidth]{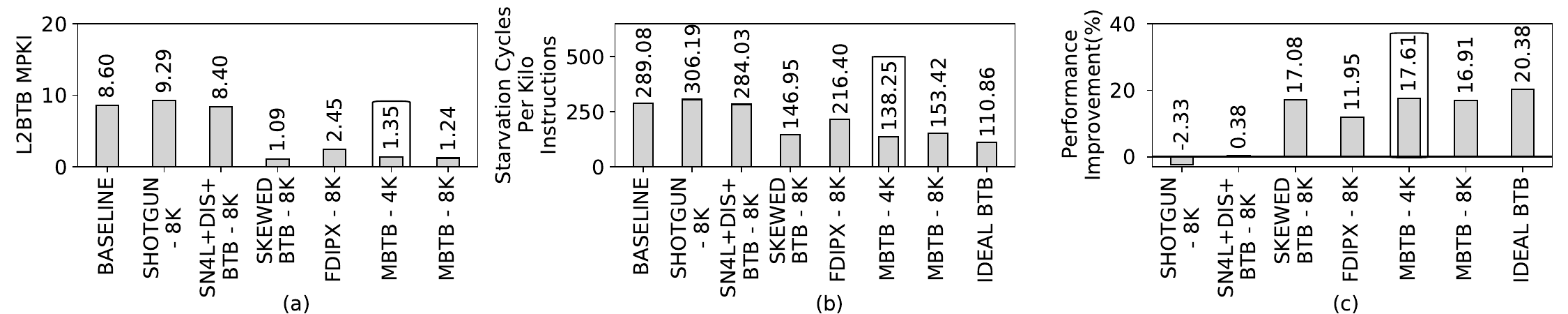}}
\caption{State-of-the-art BTB designs and MBTB evaluated on the following metrics :- (a) L2BTB Misses per Kilo Instructions (Lower the better) (b) Stall Cycles per Kilo Instructions (Lower the better) (c) Performance Improvement (Higher the better).} 
\label{fig:btb_mpki_sota}
\iftoggle{SRCSUBMISSION}{\vspace{-0.2in}}{
\vspace{-0.22in}}
\end{figure*}}

We evaluate BTB designs based on the following metrics: (i) performance in terms of Instructions per Cycle (IPC), (ii) BTB miss coverage, and (iii) SCKI. 
We first evaluate the MBTB's effectiveness in eliminating BTB misses. We then discuss the significance of skewed indexing in MBTB. Finally, we show the sensitivity studies on the number of variants, MBTB size, MBTB ways, and MBTB access latency.

\noindent \textbf{Workloads and infrastructure:}
We use 100 server workloads released publicly with CVP-1 \cite{cvp} (co-located with ISCA '18) which shows the highest sensitivity to BTB and L1I. The CVP-1 traces are from the Qualcomm server team. We use an in-house extension of ChampSim \cite{champsim} simulator. We extend ChampSim with detailed \emph{decoupled} front-end \cite{fdip}, back-end, memory hierarchy, and a detailed memory system. Our extension supports BTBs, RAS, FTQ, VIPT L1 caches, and a detailed virtual memory system (five-level page table walker (PTW) and four memory management unit (MMU) caches to back the PTW). The three-level cache hierarchy also stores the page translations. Table \ref{table:parameters} shows the parameters used, which are similar to the recent Intel Sunny Cove \cite{sunny}\cite{fdipispass}. For each trace, we first warmup for 50M instructions and then report performance for the next 50M instructions.

\noindent \textbf{BTB designs:}
Based on the performance improvement and storage overhead, we use the following recent works for our evaluation: (i) Shotgun \cite{shotgun} (ii) SN4L+Dis+BTB \cite{divide}, (iii) Skewed BTB \cite{skewseznec} and (iv) FDIP-X \cite{fdipx}.  Table \ref{table:designs} provides the details about these techniques.

\begin{figure*}[t]
    \begin{subfigure}[b]{0.65\textwidth}
                \includegraphics[width=\linewidth]{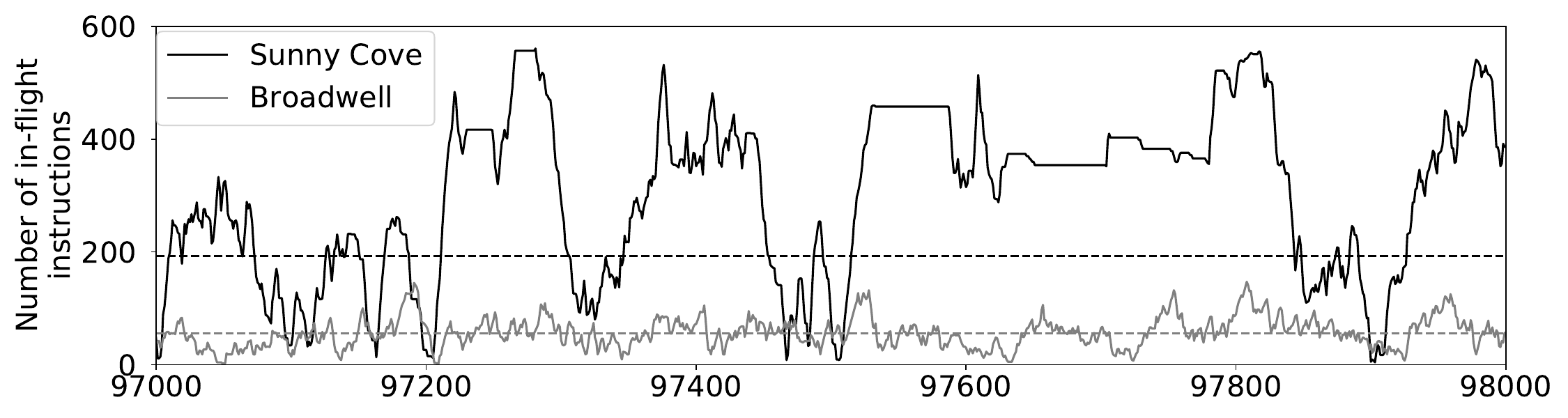}
                \caption{}
        \end{subfigure}%
        \hfill
        \begin{subfigure}[b]{0.35\textwidth}
                 \includegraphics[width=\linewidth]{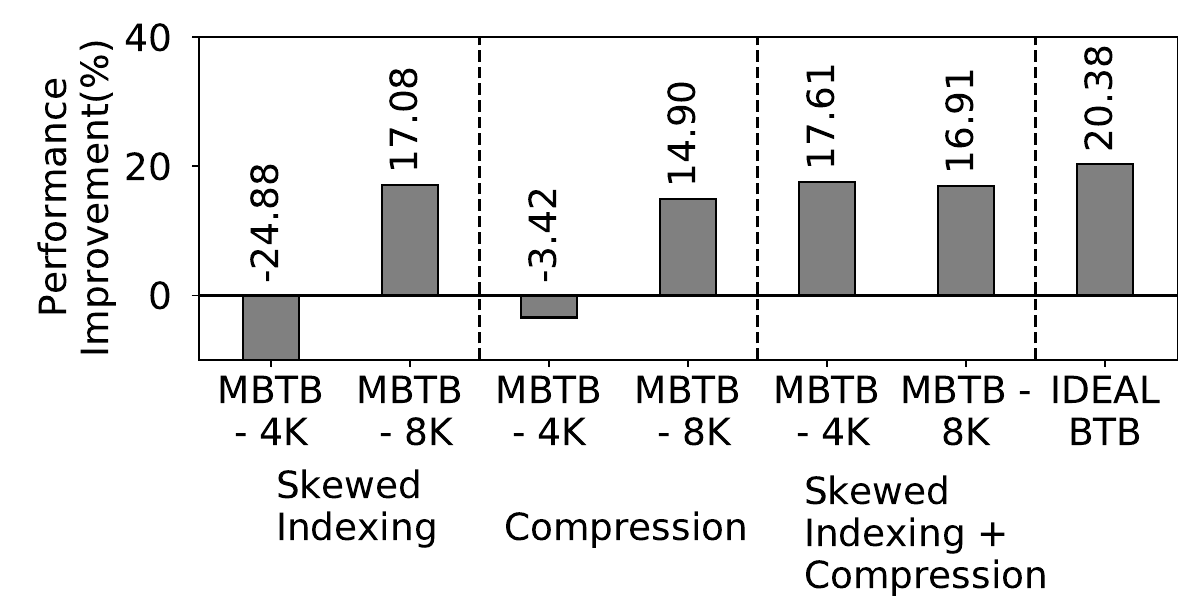}
                \caption{}
        \end{subfigure}%
\caption{(a) Number of in-flight instructions with a system similar to Intel Sunny Cove\cite{sunny} and Broadwell\cite{broadwell} micro-architecture (b) Performance improvement of MBTB with only skewed indexing, only compression and both skewed indexing and compression compared to 8K-entry L2BTB baseline.}
\label{fig:fetch_retire_diff}
\iftoggle{SRCSUBMISSION}{\vspace{-0.2in}}{
\vspace{-0.22in}}
\end{figure*}

\subsection{BTB MPKI reduction and Performance Improvement}

Figure \ref{fig:btb_mpki_sota}(a) shows the BTB misses per kilo instructions (MPKI) of different state-of-the-art BTB designs. Figure \ref{fig:btb_mpki_sota}(b) shows that the change in BTB MPKI directly correlates with the SCKI for different BTB designs. Figure \ref{fig:btb_mpki_sota}(c) shows the performance improvement with state-of-the-art BTB designs and a 4K and 8K entry MBTB. A 4K-entry MBTB compared to an 8K-entry L2BTB baseline provides an average performance improvement of 17.61\%, whereas an 8K-entry MBTB provides an average performance improvement of 16.91\%. Ideal BTB provides an average performance improvement of 20.38\%.

\textbf{MBTB} of 4K-entry decreases the BTB MPKI from 8.6 to 1.35. The decrease in BTB MPKI results in SCKI reduction from 289 to 138, which is closer to SCKI of the ideal BTB of 110. The decrease in BTB MPKI stalls the decode stage for fewer cycles, resulting in the reduction in the SCKI. MBTB of 4K-entry is able to capture, on average, 6266 branches in the L2BTB. This increase in the number of branches is what helps MBTB to reduce the BTB MPKI significantly. \emph{An 8K-entry MBTB reduces the MPKI from 1.35 to 1.24 compared to a 4K-entry MBTB, but it still provides lower performance because of an increase in access time of the MBTB. A 4K-entry MBTB can be accessed in two cycles, whereas an 8K-entry MBTB takes three cycles to access.} An 8K-entry MBTB will be more effective with higher code footprint workloads or when workloads are running in parallel on a single core \emph{i.e.} simultaneous multi-threading (SMT). In SMT, multiple threads will have a high branch footprint, and the compression scheme with MBTB will be able to hold these branches.

\textbf{Shotgun} BTB of 8K entry increases MPKI from 8.6 to 9.29 compared to a baseline system. Shotgun has a large BTB for unconditional branches and a small BTB for conditional branches. It uses BTB prefetching to prefetch conditional branches and uses the retire order stream to learn what to prefetch. Retire order stream provides better learning as it is free from the wrong execution path taken by the processor due to inaccurate branch prediction or a BTB miss. However, with modern-day high-performing processors becoming deeper and wider with each generation, these schemes fail to improve performance. For example, Intel's Broadwell\cite{broadwell} has a 192-entry Reorder Buffer (ROB) with an instruction throughput of four instructions per cycle, which increases to 352-entry ROB and six instructions per cycle with Intel's Sunny Cove processor\cite{sunnycove_anandtech}. The increase in ROB size increases the number of in-flight instructions. Figure \ref{fig:fetch_retire_diff}(a) shows the number of in-flight instructions in the pipeline for a particular phase of a server trace for a system similar to Broadwell and Sunny Cove processors. The number of in-flight instructions is taken every ten cycles during the simulation phase. The in-flight instruction includes the instruction in the fetch, decode and execute stage of the pipeline.

In the baseline system with 24 entry (192 instruction) FTQ, 60-entry decode queue, and 352-entry ROB, there can be a maximum of 604 instructions in the pipeline if all the queues are full. Figure \ref{fig:fetch_retire_diff}(a) shows that on average, in the baseline system, there are 200 instructions in-flight for this trace, with a maximum of 557. Shotgun\cite{shotgun} uses a configuration similar to the Broadwell processor in which the average number of in-flight instruction is 45, with a maximum reaching around 100. This 4x increase in the number of instructions in the pipeline leads to an increase in BTB MPKI with Shotgun. As the number of instructions increases, the fetch stage is much further in the program execution order than the retire stage, and since, Shotgun learns based on the retire order stream, it gets delayed. Shotgun is not able to issue timely prefetch requests for BTB prefetching, which results in higher BTB MPKI.

\textbf{SN4L+Dis+BTB} decreases the BTB MPKI slightly from 8.6 to 8.4. This is because it uses BTB pre-decoding to decode branches from the instruction cache blocks brought by the SN4L+Dis prefetcher. Due to large code footprint server workloads, the baseline system has an L1I MPKI of 54. SN4L+Dis prefetcher reduces the L1I MPKI to 46. Since the L1I prefetcher cannot reduce the L1I MPKI significantly, BTB pre-decoding dependent on the L1I prefetcher cannot decode the branches in a timely manner.

\subsection{Importance of Skewed Indexing}

Figure \ref{fig:fetch_retire_diff}(b) shows the performance improvement of a 4K and 8K entry MBTB with skewed indexing, compression, and both skewed indexing and compression. MBTB with only skewed indexing is same as skewed BTB. For a 4K-entry MBTB with just skewed indexing, the performance decreases by 24.88\%, whereas with compression, the performance decreases by 3.42\%. When skewed indexing and compression are used, MBTB of 4K entry improves performance by 17.61\% compared to 8K-entry baseline BTB. 

The increase in performance is mainly due to the increase in the number of branches stored in the BTB. Only skewed indexing is able to capture 4090 branches, whereas only compression captures 5496 branches. With compression and skewed indexing, MBTB is able to capture 6266 branches, which results in improved performance. The skewed indexing allows for better utilization of BTB entries, and it spreads the branches to multiple sets. This improves the compression scheme as many branches can map to entries which would not have been possible without skewed indexing.

\iftoggle{SRCSUBMISSION}{
\begin{figure}[t]
\centerline{\includegraphics[width=\linewidth]{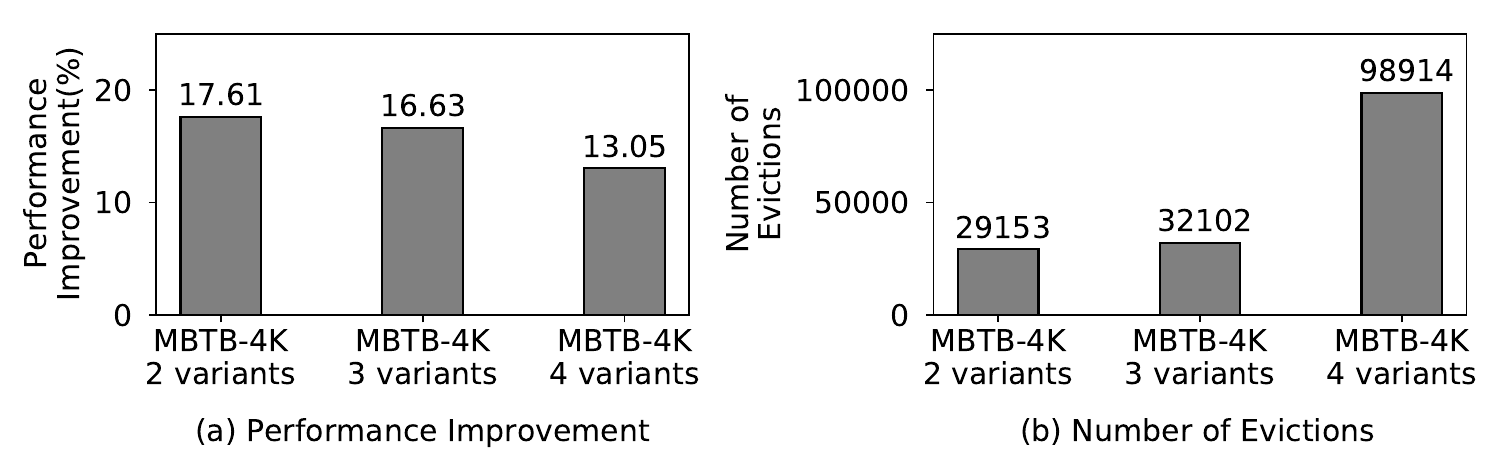}}
\caption{Variant sensitivity of a 4K entry MBTB.}
\label{fig:variant_sensitivity}
\iftoggle{SRCSUBMISSION}{\vspace{-0.2in}}{
\vspace{-0.2in}}
\end{figure}

\begin{figure*}[t]
\centerline{\includegraphics[width=0.8\linewidth]{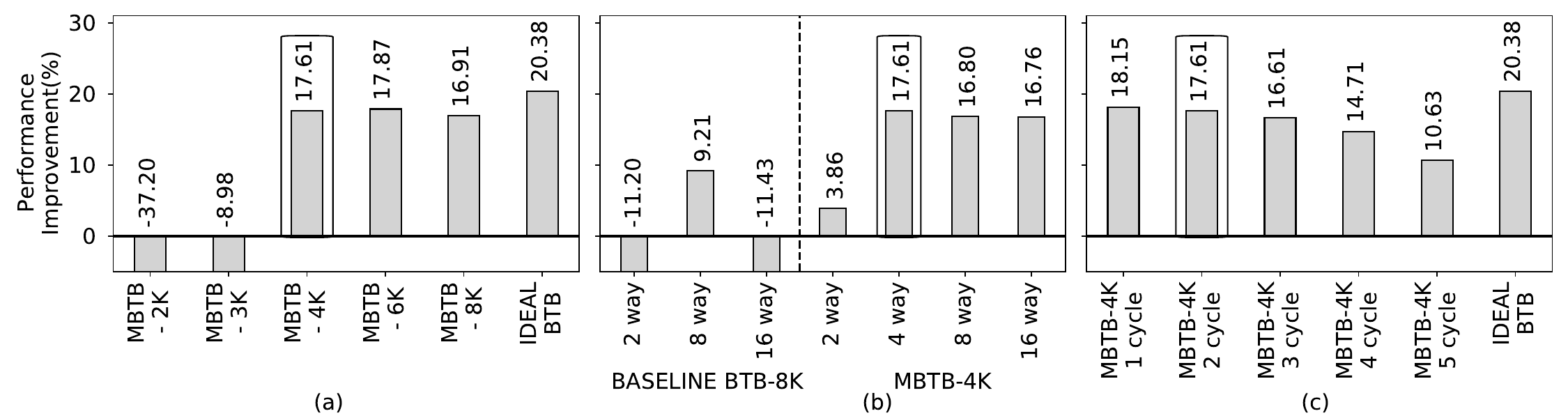}}
\caption{(a) MBTB storage sensitivity. Performance drops after 6K-entry MBTB because of increase in BTB access time (b) Way sensitivity with 8K-entry Baseline BTB and 4K-entry MBTB (c) Access latency sensitivity with 4K-entry MBTB.}
\label{fig:sensitivity_study}
\iftoggle{SRCSUBMISSION}{\vspace{-0.2in}}{
\vspace{-0.22in}}
\end{figure*}
}{}

With an 8K-entry BTB size, we see an opposite trend, and only skewed indexing performs better compared to using compression and with both skewed indexing and compression. This is because of the additional cycle latency added due to tag comparison when using compression. The increase in latency decreases the performance slightly from 17.08\% with only skewed indexing to 16.91\% with skewed indexing and compression.

\iftoggle{SRCSUBMISSION}{}{
\begin{figure}[t]
\centerline{\includegraphics[width=\linewidth]{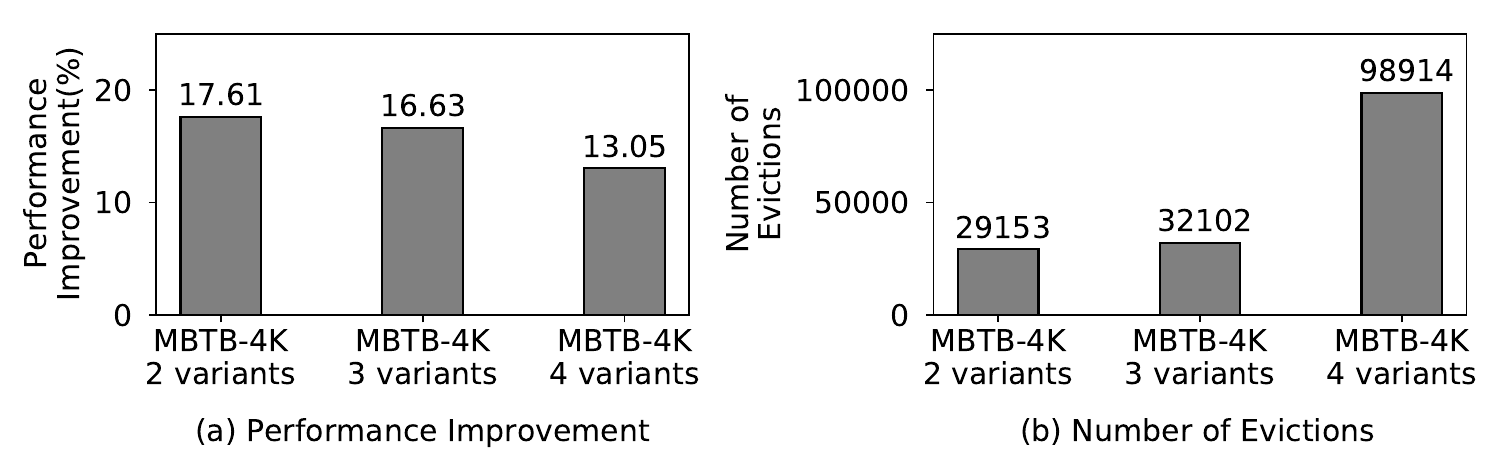}}
\caption{Variant sensitivity of a 4K entry MBTB.}
\label{fig:variant_sensitivity}
\iftoggle{SRCSUBMISSION}{}{
\vspace{-0.2in}}
\end{figure}

\begin{figure*}[t]
\centerline{\includegraphics[width=0.8\linewidth]{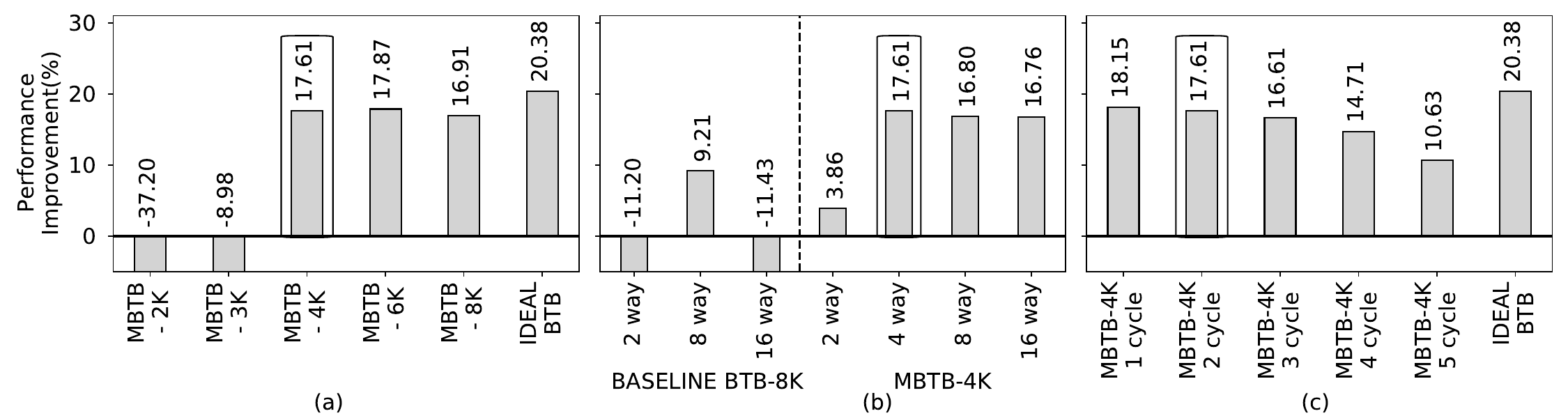}}
\caption{(a) MBTB storage sensitivity. Performance drops after 6K-entry MBTB because of increase in BTB access time (b) Way sensitivity with 8K-entry Baseline BTB and 4K-entry MBTB (c) Access latency sensitivity with 4K-entry MBTB.}
\label{fig:sensitivity_study}
\iftoggle{SRCSUBMISSION}{}{
\vspace{-0.22in}}
\end{figure*}
}

\subsection{Sensitivity study}

\noindent \textbf{Sensitivity to the number of variants:} Figure \ref{fig:variant_sensitivity}(a) shows the performance improvement of a 4K-entry MBTB with increasing the number of variants from two to four. The third variant stores four branch instructions per entry. MBTB stores branch instructions in the third variant if the offset requires less than eight bits. The fourth variant can store eight branch instructions per entry. MBTB stores branch instructions in the fourth variant if the offset requires less than four bits.

An increase in the number of variants increases the complexity of the MBTB, and it requires multiplexers and other logic gates. The increase in complexity adds an additional cycle latency for MBTB with three and four variants. RAS provides the target for return instructions, so we store the return instructions with the variant that requires the least storage per branch per entry. Three variants MBTB uses the third variant, whereas four variants MBTB uses the fourth variant.

Figure \ref{fig:variant_sensitivity}(a) shows that with increasing the number of variants, the performance decreases. This is due to an increase in the number of evictions with the increase in the number of variants as shown in Figure ~\ref{fig:variant_sensitivity}(b). In an MBTB with four variants, a branch that requires variant-1 can evict an entry with variant-3, resulting in evictions of multiple branches. Insertion of these evicted branches on reuse will evict more branches, resulting in higher BTB MPKI.

\noindent \textbf{Storage sensitivity:} Figure \ref{fig:sensitivity_study}(a) shows the MBTB performance with different storage size from 2K to 8K entries. We fix the number of ways to four and increase the number of sets. 4K-entry MBTB provides a sweet spot in terms of storage savings and performance improvement. Adding 2K-entry on top of a 4K-entry MBTB improves performance by only 0.26\%.

\noindent \textbf{Sensitivity to the number of ways:} Figure \ref{fig:sensitivity_study}(b) shows the performance improvement of 8K-entry baseline BTB and 4K-entry MBTB from 2 to 16 ways normalized to an 8K-entry baseline BTB with four ways. We observe an increase in performance when doubling the ways from four to eight in baseline BTB, but a decrease in performance with 16 ways due to an increase in access latency. 4K-entry MBTB with four ways outperforms baseline BTB with different number of ways. \\
\indent With a 2-way MBTB, the performance is less compared to a 4-way MBTB. This is because the number of branches captured reduces from 6266 branches with a 4-way MBTB to 5549 branches with a 2-way MBTB. With a 2-way MBTB, the skewed indexing can map branches to two sets instead of four sets with a 4-way MBTB. The reduction in the spread of branches cause a decrease in the number of branch instructions captured, resulting in lower performance. 

\noindent \textbf{Sensitivity to L2BTB access latency:} Figure \ref{fig:sensitivity_study}(c) shows the performance improvement of a 4K-entry MBTB when varying the access latency from one to five cycles. With a 4-cycle MBTB of 4K entry, the performance still improves by 14.71\% compared to an 8K-entry baseline BTB. This shows that as long as the BTB is accurate in providing branch targets, a slight increase in BTB latency does not impact much on the performance. This opens up for different BTB implementations with slightly slower SRAMs, which are more energy-efficient. 

\section{Summary}

In this paper, we make a case for a compressed and skewed indexed L2BTB design called MicroBTB(MBTB) to mitigate BTB bottlenecks and provide storage savings. Compression helps in increasing the storage density, whereas skewed indexing helps in providing uniform BTB access across BTB sets. Contrary to the industry trend of large BTBs, we make a case for relatively small yet effective L2BTB.

Averaged across 100 industry-provided server workloads, a 4K-entry MBTB provides 17.61\% performance improvement compared to an 8K-entry baseline BTB providing a storage savings of 47.5KB. Overall, MBTB is a lightweight and high-performance BTB design that can scale to future workloads with a higher branch footprint and can be easily adopted by the industry.

\iftoggle{SRCSUBMISSION}{
\section{ACKNOWLEDGEMENTS} 

We would like to thank Niranjan, Anuj, Neelu, Vasudha and Yashika for their helpful feedback and suggestions on the initial draft. This work is supported by the SRC grant SRC-2922.001.}{}

\bibliographystyle{ieeetr}
\bibliography{refs}

\begin{thebibliography}{10}

\bibitem{perceptron}
Tarjan and Skadron, ``{Merging Path and Gshare Indexing in Perceptron Branch
  Prediction},'' {\em TACO'05}.

\bibitem{tage}
Seznec, ``{A case for (partially)-tagged geometric history length
  predictors},'' in {\em JILP'06}.

\bibitem{shotgun}
Kumar {\em et~al.}, ``{Blasting through the Front-End Bottleneck with
  Shotgun},''

\bibitem{divide}
Ansari {\em et~al.}, ``{Divide and Conquer Frontend Bottleneck},'' in {\em
  ISCA'20}.

\bibitem{arm}
Pellegrini {\em et~al.}, ``The arm neoverse n1 platform: Building blocks for
  the next-gen cloud-to-edge infrastructure soc,'' {\em Micro'20}.

\bibitem{amd}
Suggs {\em et~al.}, ``The amd “zen 2” processor,'' {\em IEEE Micro}, 2020.

\bibitem{samsung}
Grayson {\em et~al.}, ``Evolution of the samsung exynos {CPU}
  microarchitecture,'' in {\em ISCA'20}.

\bibitem{ibmz15}
Adiga {\em et~al.}, ``{The {IBM} z15 High Frequency Mainframe Branch Predictor
  Industrial Product},'' in {\em ISCA'20}.

\bibitem{tech_scaling}
Brooks, ``{What’s the future of technology scaling?},'' 2018.

\bibitem{indirection}
Seznec, ``{Don't use the page number, but a pointer to it},'' in {\em ISCA'96}.

\bibitem{fdipx}
Asheim {\em et~al.}, ``{Fetch-Directed Instruction Prefetching Revisited},''
  2020.

\bibitem{skewseznec}
Seznec, ``{A case For Two-way Skewed-associative Caches},'' in {\em ISCA'93}.

\bibitem{sunny}
{Intel Sunny Cove.
  https://en.wikichip.org/wiki/intel/microarchitectures/sunny\_cove}.

\bibitem{shift}
Kaynak {\em et~al.}, ``Shift: Shared history instruction fetch for lean-core
  server processors,'' MICRO'13.

\bibitem{phantom}
Burcea and Moshovos, ``Phantom-btb: a virtualized branch target buffer
  design,'' in {\em ASPLOS'09}.

\bibitem{confluence}
Kaynak {\em et~al.}, ``{Confluence: Unified instruction supply for scale-out
  servers},'' in {\em MICRO'15}.

\bibitem{fdip}
Reinman {\em et~al.}, ``{Fetch Directed Instruction Prefetching},'' MICRO'99.

\bibitem{fiveptw}
{Intel 5-level Paging and 5-level EPT. https://software.intel.com/
  sites/default/files/managed/2b/80/5-level paging white paper.pdf}.

\bibitem{cvp}
{Championship Value Prediction. https://www.microarch.org/cvp1/}.

\bibitem{ipc1}
{First Instruction Prefetching Championship.
  https://research.ece.ncsu.edu/ipc/}.

\bibitem{champsim}
ChampSim. https://github.com/ChampSim/ChampSim.

\bibitem{fdipispass}
Ishii {\em et~al.}, ``{Re-establishing Fetch-Directed Instruction Prefetching:
  An Industry Perspective},'' in {\em ISPASS'21}.

\bibitem{eip}
Ros and Jimborean, ``The entangling instruction prefetcher,'' in {\em CAL'20}.

\bibitem{fnlmma}
Seznec, ``The fnl+mml instruction cache prefetcher,'' in {\em IPC-1 @ ISCA'20}.

\bibitem{skewbodin}
Bodin and Seznec, ``{Skewed Associativity Enhances Performance
  Predictability},'' in {\em ISCA'95}.

\bibitem{skewedaddress}
Bodin and Seznec, ``Skewed associativity improves program performance and
  enhances predictability,'' in {\em TACO'97}.

\bibitem{fast}
Jim\'{e}nez, ``{Fast Path-Based Neural Branch Prediction},'' in {\em MICRO'03}.

\bibitem{merge}
McFarling, ``{Combining Branch Predictors},'' 1993.

\bibitem{ipcp}
Pakalapati and Panda, ``{Bouquet of instruction pointers: Instruction pointer
  classifier-based spatial hardware prefetching},'' in {\em ISCA'20}.

\bibitem{rrip}
Jaleel {\em et~al.}, ``{High Performance Cache Replacement Using Re-Reference
  Interval Prediction (RRIP)},'' in {\em ISCA'10}.

\bibitem{broadwell}
{Broadwell Micro-architecture.
  https://en.wikichip.org/wiki/intel/microarchitectures/ broadwell}.

\bibitem{sunnycove_anandtech}
{Intel Sunny Cove micro-architecture. https://www.anandtech.com/show/
  14514/examining-intels-ice-lake-microarchitecture-and-sunny-cove/3}.

\end{thebibliography}

\end{document}